\documentclass[conference]{IEEEtran}

\usepackage[utf8]{inputenc}
\usepackage[T1]{fontenc}
\usepackage[english]{babel}
\usepackage{array}
\usepackage{shortvrb}
\usepackage{listings}
\usepackage{amsmath}
\usepackage{amsfonts}
\usepackage{enumerate}
\usepackage{graphicx}             
\usepackage[caption=false]{subfig}
\usepackage{subfloat}
\usepackage{alltt}
\usepackage{url}
\usepackage{xspace}
\usepackage{indentfirst}
\usepackage{eurosym}
\usepackage{wasysym}
\usepackage{color}
\usepackage{varwidth}
\usepackage{framed}
\usepackage{tcolorbox}
\usepackage{authblk}
\tcbuselibrary{theorems}
\usepackage{xifthen}
\usepackage{multirow}
\usepackage{enumitem}
\usepackage{tikz}
\usetikzlibrary{positioning,decorations.pathreplacing,calc,fit,shapes.geometric, arrows,shapes.symbols,arrows.meta}
\usepackage{mathtools}
\usepackage{enumitem}
\usepackage{stackrel}
\usepackage{colortbl}
\usepackage{nth}
\usepackage{booktabs}
\usepackage{balance}
\usepackage{changepage}
\usepackage{siunitx} 

\graphicspath{{.}{./Pictures/}}

\definecolor{mygreen}{rgb}{0,0.6,0}
\definecolor{mymauve}{rgb}{0.58,0,0.82}
\definecolor{mygray}{rgb}{0.5,0.5,0.5}
\definecolor{RougeLosange}{RGB}{255,42,6}
\definecolor{BleuBloc}{RGB}{0,156,253}
\definecolor{MauveBloc}{RGB}{131,54,188}
\definecolor{OrangeBloc}{RGB}{222,106,16}
\definecolor{MACRouge}{RGB}{255,38,0}
\definecolor{MACBleu}{RGB}{4,51,255}
\definecolor{MACVert}{RGB}{0,143,0}
\definecolor{MACOrange}{RGB}{255,147,0}
\definecolor{dline}{RGB}{255,38,0}
\definecolor{done}{RGB}{4,51,255}
\definecolor{todo}{RGB}{0,143,0}
\definecolor{lowerbound}{RGB}{255,147,0}
\definecolor{upperbound}{RGB}{255,64,255}
\definecolor{variter}{RGB}{146,144,0}
\definecolor{vardata}{RGB}{169,169,169}
\definecolor{varsize}{RGB}{148,23,81}
\definecolor{unmodified}{RGB}{0,150,255}
\definecolor{green-yellow}{rgb}{0.68, 1.0, 0.18}
\definecolor{red0946}{RGB}{255,38,0}
\definecolor{green0946}{RGB}{43,147,43}
\definecolor{Sepia}{HTML}{671800}

\newcommand{\dfn}[1]{\textit{#1}}            

\newcommand{\traceroute}{\texttt{traceroute}\xspace}

\newcommand{\ping}{\texttt{ping}\xspace}
\newcommand{\tnt}{\texttt{TNT}\xspace}
\newcommand{\scamper}{\texttt{scamper}\xspace}
\newcommand{\utnt}{$\upsilon$\tnt}
\newcommand{\sctracediff}{\texttt{sc\_tracediff}\xspace}
\newcommand{\musl}{\texttt{musl}\xspace}
\newcommand{\lwip}{\texttt{lwip}\xspace}

\begin{document}

\title{\utnt: Unikernels for Efficient and Flexible Internet Probing}
\author[1]{Maxime Letemple}
\author[2]{Gaulthier Gain}
\author[2]{Sami Ben Mariem}
\author[2]{Laurent Mathy}
\author[2]{Benoit Donnet}
\affil[1]{Institut Polytechnique de Bordeaux, France}
\affil[ ]{\texttt{mletemple@ipb.fr}}
\affil[2]{Universit\'e de Li\`ege, Montefiore Institute, Belgium}
\affil[ ]{\texttt{\{firstname.name\}@uliege.be}}
\date{}
\renewcommand\Affilfont{\itshape\small}

\maketitle

\begin{abstract}

The last twenty years have seen the development and popularity of
network measurement infrastructures. Internet measurement platforms
have become common and have demonstrated their relevance in Internet
understanding and security observation. However, despite their
popularity, those platforms lack of flexibility and reactivity, as
they are usually used for longitudinal measurements. As a consequence,
they may miss detecting events that are security or Internet-related.
During the same period, operating systems have evolved to virtual
machines (VMs) as self-contained units for running applications, with
the recent rise of unikernels, ultra-lightweight VMs tailored for
specific applications, eliminating the need for a host OS.

In this paper, we advocate that measurement infrastructures could take
advantage of unikernels to become more flexible and efficient. We
propose \utnt, a proof-of-concept unikernel-based implementation of
\tnt, a \traceroute extension able to reveal MPLS tunnels. This paper
documents the full toolchain for porting \tnt into a unikernel and
evaluates \utnt performance with respect to more traditional
approaches. The paper also discusses a use case in which \utnt could
find a suitable usage. \utnt source code will be released upon paper
acceptance.

\end{abstract}

\section{Introduction}\label{intro}
For more than twenty years now, Internet measurement platforms have
become common~\cite{measurementPlatforms}. Most of them have been
deployed by researchers and came with predefined measurement
capacities~\cite{skitter,ark,ripe,mlab,iplane}.  Measurements are
usually performed by platform owner and collected data set is often
made available to the research community. Those platforms typically
rely on dedicated hardware, located in well-known places or run by an
army of volunteers. In parallel to those academic platforms, several
measurement infrastructures~\cite{dimes,neti,dipzoom} have been
deployed following the SETI@home~\cite{seti} model.  That kind of
infrastructure comes with the advantage of potentially increasing the
probing sources, but at the cost of development difficulties
(tools must support a variety of hardware, operating systems, and
local configuration) and irregular data collection.  Further, several
consortiums~\cite{planetlab,edgenet,scriptroute} have also built
distributed platforms on which researchers can possibly deploy
their own network measurement experiences.

All those solutions require either dedicated hardware (e.g., RIPE
Atlas~\cite{ripe}, Archipelago~\cite{ark}, or even
PlanetLab~\cite{planetlab}), either an adaptation of tools to running
operating system and libraries (e.g., PlanetLab~\cite{planetlab} or
Dimes~\cite{dimes}). More recent ones rely on Kubernetes and VM
deployment (e.g., EdgeNet~\cite{edgenet}).

Unfortunately, none of those solutions is flexible and easily
deployable. As such, they are not built to quickly react to events,
either related to security, network
outages~\cite{outages,measurements_security}, or even Internet
topology dynamicity discovery~\cite{bgp_guided}. By quickly, we mean
it should be instantiated on-demand (loading time must be as quick as
possible), should require the lowest memory footprint, and shutdown
when the measurement is over.

In parallel to this, the last twenty years have seen the development
of advanced operating systems. The advent of public clouds initially
relied on hardware virtualization, using virtual machines (VMs) as
self-contained units for running applications. While cost-effective,
VMs were heavyweight, consuming substantial resources and storage due
to their full OS image. This led to the adoption of containerization,
with technologies like Docker~\cite{docker_url} and
LXC~\cite{lxc_url}, which share the host OS kernel and reduce resource
usage. However, containers have security concerns due to their large
attack surface~\cite {docker-exploits,ncc-hardening}. To strike a
balance between performance and isolation, a newer paradigm emerged:
\dfn {unikernels}~\cite{unikernel}. Unikernels are ultra-lightweight
VMs tailored for specific applications, eliminating the need for a
host OS. Including only essential OS components offers improved
performance and a reduced attack surface, making them a promising
alternative to VMs and containers~\cite
{gain_socc,kantee,Kivity2014,unikraft,clothing}.

The main challenge with unikernels is the need for manual porting of
existing applications, involving intricate processes like component
extraction, API compliance, and optimization. Fortunately, frameworks
such as Unikraft~\cite{unikraft} and its tools suite have been
developed to make it easier to port existing applications to
unikernels~\cite{unikraft_tools,lefeuvre2023loupe}. The goal is to
produce unikernels with small image sizes, quick boot times, and
minimal memory usage, simplifying development and deployment.

This paper advocates that unikernels can be the foundation of new,
flexible, and efficient measurement infrastructures. In particular,
this paper introduces \utnt, a proof-of-concept unikernel-based
implementation of \tnt, a \traceroute extension able to reveal MPLS
tunnels and to provide hardware vendors pieces of information~\cite
{tnt-tma,tnt-tnsm}. Our paper makes the following contributions: ($i$)
It describes the \utnt implementation, serving as a first example of
the full toolchain for porting any network measurements software into
a unikernel. ($ii$) It evaluates \utnt performance, demonstrating that
it is more flexible than traditional implementation as it consumes
less CPU and memory. ($iii$) It discusses a use case in which the
flexibility of unikernel based probing tool, such as \utnt, could find
a suitable usage. We also evaluate its performance in such a context,
compared to more traditional approaches.

The remainder of this paper is organized as follows: Sec.~\ref
{background} provides the required background for the paper;
Sec.~\ref{utnt} discusses how we implemented \utnt and Sec.~\ref
{evaluation}  evaluates its performance; Sec.~\ref
{usecase} introduces a use case in which \utnt would find a suitable
usage; Sec.~\ref{discussion} discusses the limits of our work;
finally, Sec.~\ref{ccl} concludes this paper by summarizing its main
achievements.
\section{Background}\label{background}
This section introduces the required background for this paper. In
particular, Sec.~\ref{background.unikernels} discusses unikernels in
general. Sec.~\ref{background.unikraft} presents
\dfn{Unikraft}, a framework for porting applications on unikernels.
Finally, Sec.~\ref{background.memory_deduplication}, gives an
overview of memory deduplication, a technique for merging identical
memory pages.

\subsection{Unikernels}\label{background.unikernels}
\begin{figure}[!ht]
  \centering
  \includegraphics[width=\linewidth]{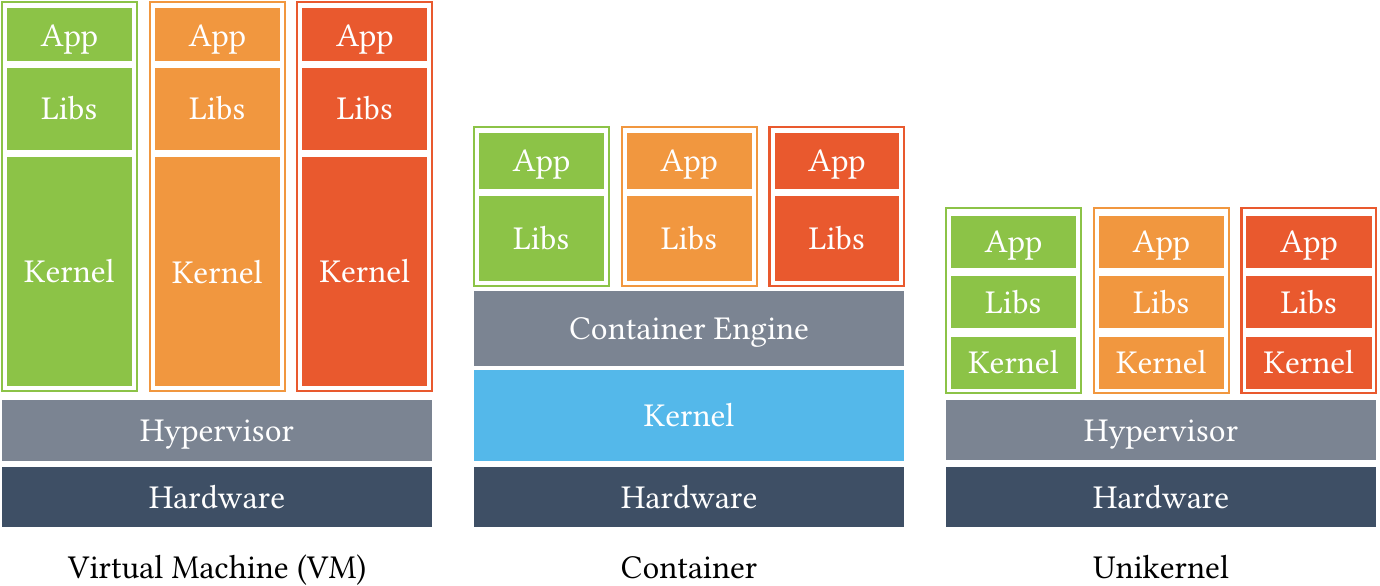}
  \caption{Comparison of a virtual machine (VM), a container, and a unikernel.}
  \label{fig:containers-unikernels-vms.pdf}
\end{figure}

The concept of a unikernel represents a recent innovation where an
application is tightly integrated with the underlying kernel. In
simpler terms, software is compiled to include only the necessary OS
functionality, such as required system calls and drivers, forming a
single statically-compiled executable image. This single address space
design means that unikernels do not maintain separate user and kernel
address spaces, and all threads, as well as the kernel, share the same
page table. The primary advantage of this approach lies in its
capacity to minimize the potential attack surface and the
exploitability of the operating system code. Unlike virtual machines
and containers, which often come bundled with an excess of tools and
libraries beyond what the running application truly needs, unikernels
exclusively contain essential operating system functions.
Fig.~\ref{fig:containers-unikernels-vms.pdf} gives a high-overview of
the different paradigms between VMs, containers and unikernels.

In addition to enhancing security, unikernels also offer performance
improvements. They utilize a single address space without
distinguishing between kernel-space and user-space. Consequently,
system calls are akin to regular function calls, sidestepping the
performance overhead associated with context switches and data
transfers between user and kernel spaces.

Unikernels can be categorized into two primary types: ($i$) \dfn
{POSIX-compliant unikernels}. These unikernels have the capability to
execute both existing and legacy applications through the use of
cross-compiling techniques. Typically, they are constructed around a
customized kernel and exhibit a larger code base due to their
increased resource requirements. Nonetheless, these platforms offer a
straightforward approach to migrate traditional software, originally
designed for virtual machines and containers, into the unikernel
environment by simply requiring recompilation. Unikraft~\cite
{unikraft} is an example of this type of system. It is tailored to
run unaltered Linux applications on the KVM (Kernel
Virtual Machine~\cite{kvm}) hypervisor. Similarly,
within this classification, Rumprun~\cite{kantee} provides reusable
kernel-quality components that facilitate the creation of highly
customized unikernel images with minimal overhead. ($ii$)  \dfn
{Language-based unikernels}. In contrast, this second category
pertains to the development of minimalist operating systems with a
custom API. Unlike the previous approach, this model does not aim to
optimize existing code; instead, it concentrates on providing a set
of tools for the rapid assembly of new components, eliminating the
need to address underlying services such as memory allocators and
drivers. However, this concept does have the drawback of generating a
code base that is generally incompatible with existing applications.
Consequently, it necessitates the rewriting of legacy code to conform
to the defined platform's API. MirageOS~\cite{unikernel}, developed
using OCaml, represents an example of this architecture. It serves as
a complete, ground-up set of protocol libraries designed for the
construction of specialized unikernels that operate on the Xen
hypervisor~\cite{xen_barham_2003}.

\subsection{Unikraft}\label{background.unikraft}

While traditional OS development is typically divided between
monolithic kernels, where all critical system functions like device
drivers are closely integrated into a single
kernel~\cite{Jacobsen2015,linuxKernel}, and modular micro-kernels that
prioritize isolating OS
components~\cite{microkernel_perf2,microkernel_perf}, Unikraft stands
out by adopting a unique approach. It combines the monolithic design
(no protection between components) with the modularity provided by
micro-kernels. Unikraft leverages modularity to enable specialization,
dividing OS functionality into distinct components that only interact
through well-defined API boundaries. Rather than compromising API
boundaries for the sake of performance, Unikraft ensures performance
through meticulous API design and static linking. To fulfill the
overarching principle of modularity, Unikraft relies on two primary
components.

\textbf{($i$) Micro-libraries and pools}: Micro-libraries, (micro-libs
for short), are software components which implement one of the core
Unikraft APIs. Within a given pool of libraries, each of them adheres
to the same API standard, making them fully interchangeable.
Furthermore, micro-libraries can integrate functionality from external
library projects such as OpenSSL~\cite {openssl}, \musl~\cite{musl},
Protobuf~\cite{protobuf}, and more. They can also incorporate features
from applications like SQLite~\cite{sqlite},
Nginx~\cite{nginx}, or even adapt to different platforms like
Solo5~\cite{solo5}, Firecracker~\cite {firecracker2020}, or Raspberry
Pi 3 and architectures (e.g., x86, ARM, etc.). The size of
micro-libraries can vary widely, ranging from extremely compact ones,
such as those containing basic boot code (e.g.,
\texttt{ukboot}~\cite{ukboot}), to substantial ones, like those
providing comprehensive libc (e.g., musl~\cite{musl}) or network
support (e.g., \lwip~\cite{lwip}). The Unikraft project maintains
several micro-libraries available on Github~\cite{unikraft_github},
but users and developers can also create their own.

\begin{figure}[!ht]
  \centering
  \includegraphics[width=\linewidth]{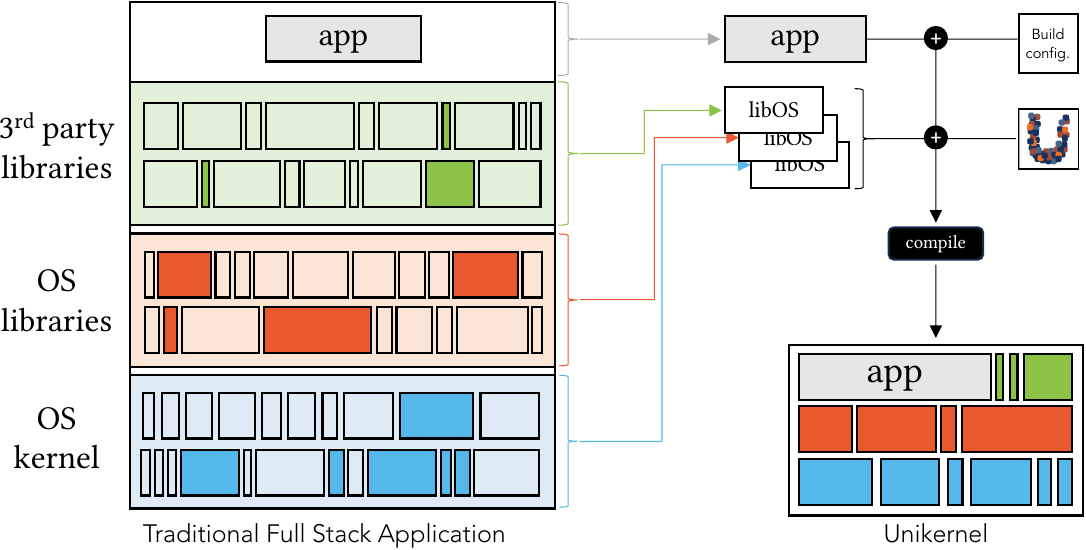}
  \caption{From a traditional full stack application to an optimized
  Unikraft-based unikernel. The application to be ported
  is shipped with the essential libraries and OS components into a
  unikernel.}
  \label{fig:unikernel_unikraft}
\end{figure}

\textbf{($ii$) Buildsystem}: Unikraft provides a user-friendly menu
based on KConfig\footnote{Kconfig is a configuration system
providing a hierarchical menu-driven interface for
enabling/disabling features.}, allowing users to choose the
micro-libraries they want to incorporate into their application
during the build process. Users can also specify the platform(s) and
CPU architectures to target and configure individual micro-libs if
desired (i.e. memory allocators, network protocols support, etc).
The build system efficiently handles the compilation and linking of
all chosen micro-libs, resulting in the creation of one binary per
selected platform and architecture. Fig.~\ref{fig:unikernel_unikraft} 
shows how a unikernel can be built using Unikraft.

To port an existing application to Unikraft, developers need to
provide three essential files: ($i$) A \texttt{Makefile} file
designating the Unikraft repository's location relative to the
application's repository, as well as any external library
repositories. ($ii$) A \texttt{Makefile.uk} file which serves as the
principal \texttt{Makefile} for specifying sources to build, including
paths, flags, and application-specific targets. ($iii$) A
\texttt{Config.uk} file, which populates Unikraft's menu with
application-specific configuration options. In most cases, porting an
existing library or application to Unikraft requires few if any
changes to the actual code: most of the work consists of creating a
Unikraft makefile called \texttt{Makefile.uk} that Unikraft uses to
compile the source code. When this step is done, some changes in the
application code may be applied to avoid runtime errors (e.g.,
stubbing fork, exec, etc.).

\subsection{Memory Deduplication in Virtualized Environments}\label{background.memory_deduplication}
Running multiple VMs or unikernels can significantly inflate memory
consumption because of the inherent isolation between instances. To
mitigate excessive memory usage across various
instances~\cite{gain_socc,ArcangeliKSM}, cloud providers may employ a
memory deduplication scanner. This technology is particularly useful in
virtualized environments where multiple VMs share the same physical
hardware resources. Memory deduplication scanners analyze the contents
of a system's memory to detect identical memory pages\footnote{A page
represents a uniform, fixed-size segment of virtual memory, and is
characterized by a single entry within the page table. It is the
smallest unit of data for memory management in a virtual memory
operating system.} and merges them into a same single frame.
For instance, the \dfn{Kernel Samepage Merging}
(KSM~\cite{ArcangeliKSM}) is the default memory deduplication scanner
available in the Linux Kernel. There exist also other scanners such as
\dfn{UKSM}~\cite{NaiXiaUksm} or \dfn{KSM++}\cite{MillerKSMUI} which
deliver better performance than KSM, but these are not integrated into
the Linux kernel and require the use of custom kernels.

\section{\utnt}\label{utnt}
This section is dedicated to \utnt, the porting of \tnt (a modified
\traceroute driver for \scamper) into a unikernel using Unikraft.
Sec.~\ref{utnt.overview} gives an overview of \scamper and \tnt.
Sec.~\ref{utnt.implementation} documents how \utnt has been
implemented.

\subsection{Overview}\label{utnt.overview}

\scamper~\cite{scamper} is a modern implementation of
\traceroute~\cite{traceroute}. In addition to traditional \traceroute
(in both IPv4 and IPv6), \scamper comes with multiple tools: \ping,
alias resolution~\cite{alias}, DNS probing, and load balancing
discovery~\cite{loadBalancing}. \scamper has been designed to probe
the Internet in parallel, at a specified packet-per-second rate. This
paper considers an additional layer above \scamper: \tnt~\cite
{tnt-tma,tnt-tnsm}. \tnt uses \traceroute-like probing for revealing
the presence of MPLS tunnels~\cite{mpls-invisible}. Further, \tnt
provides, for each collected IP interface, its fingerprint in order to
identify the hardware vendor~\cite{fingerprinting}.

Scamper is based on two processes to operate: a listening
server, which waits for a command, performs the measurements and sends
back the results to the second process that saves the result into a
file. Initiating \scamper involves commencing with the primary
'\scamper' program, which waits for measurement commands.
Subsequently, in a distinct terminal, the designated command for the
desired measurement should be executed. For example, to execute a
standard \traceroute, \sctracediff is employed, while \tnt is used for
the modified traceroute (\tnt).

\subsection{Implementation}\label{utnt.implementation}

To transform \tnt into a unikernel, denoted as \utnt, we used
Unikraft~\cite{unikraft} and one of its toolsets~\cite{unikraft_tools}
tailored to streamline the porting process. We opted for the Unikraft
framework as our unikernel framework for a variety of compelling
reasons. Firstly, it boasts an extensive library and application
collection, all are open-source and available on
GitHub~\cite{unikraft_github}. Secondly, it is actively maintained and
continually evolving. Additionally, its POSIX compatibility eliminates
the need to rewrite applications from scratch. Lastly, a range of
tools is available for application porting. Nevertheless, it is worth
noting that alternative unikernel frameworks like Nanos~\cite{nanos}
and Rumprun~\cite{kantee} remain viable options.

\begin{figure}[!ht]
  \centering
  \includegraphics[width=\linewidth]{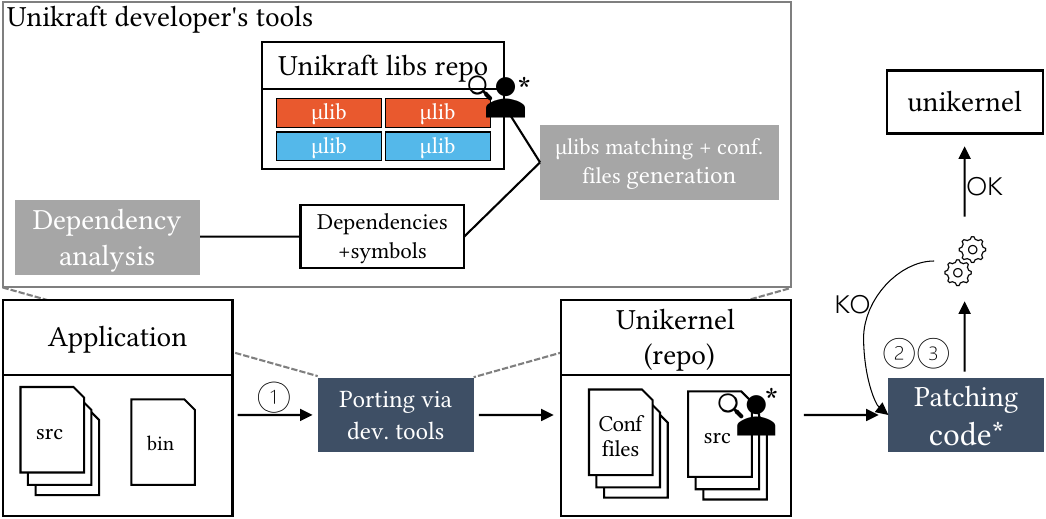}
  \caption{High-level overview of the different steps to port
  an application into a unikernel. The Unikraft developers tool is
  used to perform micro-libraries matching and to generate the required
  configuration files. After this step, the application's codebase has been
  patched to be compliant with Unikraft.}
  \label{fig:diagram_toolchain}
\end{figure}

In Fig.~\ref{fig:diagram_toolchain}, we outline the $3$ general steps
involved to port an application as unikernel: (1) Using a developer
assisting tool to perform the libraries matching and generating the
required configuration files for Unikraft. (2) Adapting the given
application's codebase to support Unikraft paradigm (e.g., single
address space and single process). (3) Updating external libraries by
modifying their codebase to be compliant with the ported application.
These last two steps can either be skipped (no compilation or linking
error) or be repeated several times until the application is
successfully built using the Unikraft build system.

We ported Scamper+\tnt according to these steps. During the first
step, we used a developer tool~\cite{unikraft_tools}, written by the
Unikraft community, which analyses the application's symbols, both in
its binary and source files, endeavoring to match them with the
symbols found in Unikraft's micro-libraries. As a result, it
automatically generates a fully configured \texttt {Makefile.uk}
containing the necessary source files and a \texttt {Makefile} with
the required micro-libraries. When applied to the \scamper folder, the
tool produced a \texttt {Makefile.uk} comprising $82$ source files and
a \texttt{Makefile} with only two external libraries: \musl as the C
library (libc) and \lwip for the networking support. Please note that
core libraries such as \texttt{uksched}, \texttt{ukboot}, etc., are
also used but not listed in the makefile; only external libraries are
included.

Throughout the second step, we undertook the task of combining
\scamper and \tnt into Unikraft. Given that Unikraft exclusively
supports single-process applications, we made some necessary code
adaptations to remove the multi-process support. This adjustment
involved two key steps: first, we modified the configuration file in
\scamper (\texttt {config.h}) to disable multi-process support by
modifying some core variables. For instance, we disabled the privilege
separation daemon, allowing to run two processes - one with and one
without privileges. Second, we updated the main file of \tnt to launch
two threads, one for \scamper and another for \tnt. Additionally, we
converted the original socket-based communication between the two
processes into multithreaded communication. Indeed, \scamper
originally used a specific sockets implementation for inter-process
communication, which was incompatible with the Unikraft framework,
which only supports a certain type of socket family, provided by the
\lwip network stack. To establish communication between \scamper and
\tnt in this context, we made changes to the core code, specifically
by adjusting some functions to enforce the use of \lwip's sockets
while preserving the rest of the code unchanged. In addition, we also
removed irrelevant code for Unikraft, such as the code related to user
permission controls (e.g., \texttt{setgroups}, etc) and root directory
management (e.g., \texttt{chroot}, etc.).

After that, we had to make a few minor adjustments to the \lwip and
\musl libraries to ensure the proper functioning of \utnt. For
instance, we modified some \lwip functions to prevent the automatic
inclusion of an IP header resulting in a conflict between \lwip and
\scamper. Additionally, we also modified specific \lwip options by
updating the \lwip configuration (e.g., respond to broadcast pings,
enable IP forwarding, etc.) to be compliant with \scamper behaviour.
For \musl, we updated the Makefile to include additional headers
required by \scamper. Once all these steps were completed, we were
able to successfully compile and build \utnt.

We give an overview of our changes in Table~\ref{table:changes}. To
count the lines of code (LoC), we used CLoc~\cite{cloc} and considered
only the C/C++ files with their associated header files (only code
without any comment). The last column gives the percentage of changes
compared to the original codebase. All the patches containing our
changes can be found with our setup in our GitHub
repository\footnote{URL redacted for anonymity.}.

{
\setlength{\doublerulesep}{0pt}
\setlength{\tabcolsep}{5pt}
\begin{table}[!ht]
  \caption{Overview of the changes made to the different libraries.}
  \centering
  \begin{tabular}{c | c | c | c}
   \hline\hline\hline\hline
   \textbf{Micro} & \textbf{\# File(s)} & \textbf{\# LoC} & \textbf{\% of} \\ 
   \textbf{Libraries} & \textbf{updated} & \textbf{updated} & \textbf{changes} \\
   \hline\hline\hline\hline
   core (\texttt{ukboot}, etc)  & / & / & / \\
   \musl (libc)    & 1 & 5 & <0.01\%\\
   \lwip (network) & 3 & 12 & 0.01\%\\
   \hline
   \scamper & 5 & 110 & 0.2\% \\
   \tnt     & 1 & 70  & 1.8\% \\ 
   \hline\hline\hline\hline
   \dfn{Total} & 11 & 197 & / \\ 
   \hline\hline\hline\hline
  \end{tabular}
  \label{table:changes}
\end{table}
}

After successfully porting the application, the next step involves
establishing an Internet connection to perform network measurements.
To achieve this, we create and setup a network bridge, which is then
linked to the \utnt via a configuration file.

\section{Performance Evaluation}\label{evaluation}

In this section, we showcase the experimental results obtained through
a comparison of \tnt across three distinct architectures. These
paradigms include: ($i$) Adapting \tnt for use within a conventional
Debian virtual machine (VM) with Vagrant for simplified management, ($ii$) Containerizing
\tnt through the use of a Docker container, and ($iii$) Porting \tnt into
a unikernel (i.e., \utnt) by using Unikraft as codebase and
Firecracker~\cite{firecracker2020} as a hypervisor with KVM (Kernel
Virtual Machine~\cite{kvm}) support. The primary objective of these
experiments is to assess \utnt's performance across various
dimensions, such as memory usage, CPU utilization, file size, and
total execution time across these distinct architectures.

\subsection{Methodology}\label{evaluation.methodo}

To conduct an equitable comparison of the various architectures, we
standardized the environment by employing identical versions of
\scamper and \tnt across all three cases. In the case of the VM, we
relied on the capabilities of Vagrant, a versatile and dynamic virtual
machine management tool. We crafted a \texttt{Vagrantfile} that
instantiated a minimal Debian server boasting $256$~MB of memory and
with \texttt{libvirt}~\cite{libvirt_url} (a convenient way to manage
VM and virtualized functionalities) as the provider. This Vagrant
environment executed a shell script responsible for running \scamper
and \tnt. For Docker, we employed the official Debian image and made
adaptations to the underlying Dockerfile to enable the execution of
\scamper and \tnt. Notably, in both setups, we set the same
destination target (i.e., 1.1.1.1) and maintained consistent parameter
settings for \tnt.

All the following experiments were performed on a Debian GNU/Linux 11
(bullseye) with a Linux kernel $5.10.162$. Firecracker version
$1.2.1$, Vagrant $2.2.14$, and Docker $20.10.5$ have been used. The
host machine used for the experiments has 32 GB of RAM and an Intel
(R) Xeon(R) CPU E5-2620v4 @2.10GHz with 16 cores. In addition,
Unikraft Pandora $0.15.0$ has been used for the following experiments.

Ultimately, we opted for KSM\footnote{Configured with the default
parameters~\cite{url_ksm}.}, the default memory deduplication scanner in
Linux, as it is directly integrated into the kernel. Using another
deduplication scanner that necessitates a custom kernel appears
less relevant, given the constraints related to underlying host kernels
in cloud platforms~\cite{url_google_cloud,url_aws}.

\subsection{Results}\label{evaluation.results}

\begin{figure}[!htp]
  \centering
  \includegraphics[width=0.95\linewidth]{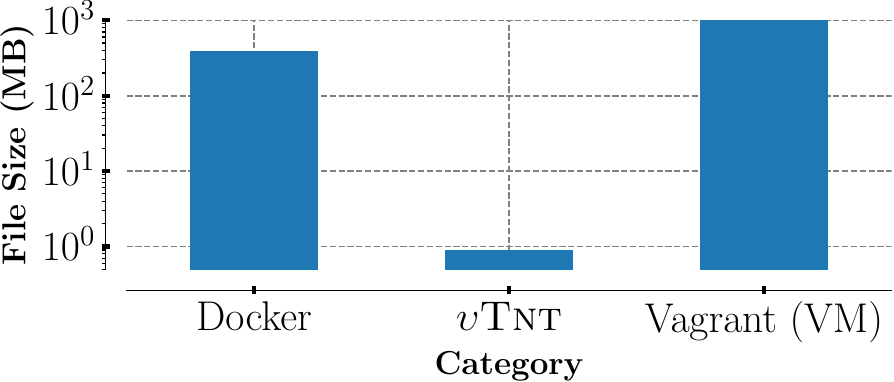}
  \caption{Size of a Docker image, a Vagrant image, and \utnt.
  Using unikernel considerably reduced disk usage.}
  \label{evaluation.results.fig.filesize}
\end{figure}

As depicted in Fig.~\ref{evaluation.results.fig.filesize}, the \utnt
unikernel has a tiny size, occupying less than $1$~MB of space. This
impressive size reduction can be attributed to Unikraft's capability
to execute \dfn{Dead Code Elimination} (DCE -- i.e., an optimization
that removes code which does not affect the program results) on the
static underlying file, efficiently removing extraneous code and
resulting in a minimal image containing only essential components. In
stark contrast, the Docker image is $430$ times larger than \utnt.
Similarly, the Vagrant image proved even more substantial, with an
image $1,170$ times larger than \utnt.

\begin{figure}[!htp]
  \centering
  \includegraphics[width=0.95\linewidth]{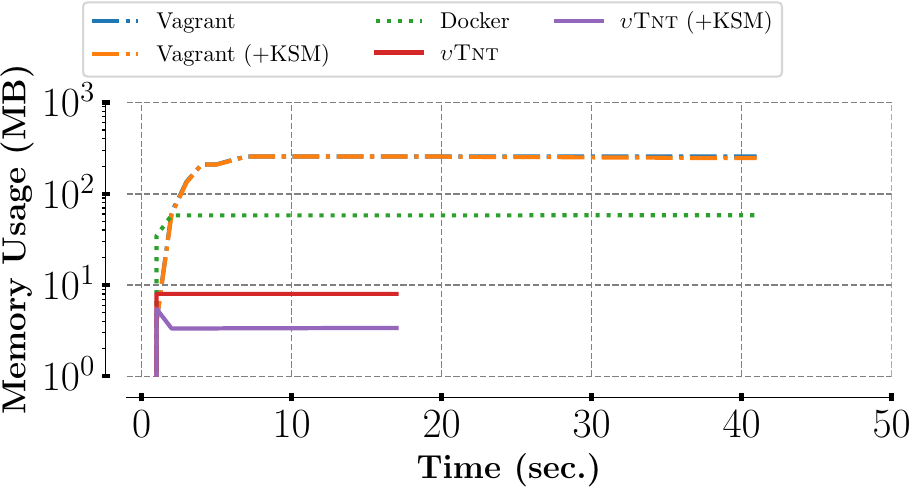}
  \caption{Memory consumption of \utnt, compared to \scamper
  and Vagrant. In addition, two additional configurations denoted
  ``(+KSM)'' are evaluated using a memory deduplication scanner to merge
  identical pages.}
  \label{evaluation.results.fig.memory}
\end{figure}

After evaluating image sizes, our next step was to measure the memory
utilization across each configuration. For each one, we measured the
overall memory footprint (hypervisor and guest operating system for
\utnt $+$ Vagrant and Docker engine $+$ container for Docker) of a
single instance. In addition to the three fundamental setups, we
introduced an additional one utilizing KSM, denoted as ``(+KSM)''. Our
initial comparison focused first on the three original configurations,
excluding KSM. As depicted in Fig.~\ref{evaluation.results.fig.memory},
our observations reveal that \utnt consumes approximately $7.2$ times
less memory than Docker and $32$ times less memory than the Vagrant
configuration. It is worth noting that the Vagrant configuration is
the most memory-intensive, given its requirement for a complete
operating system to operate. In contrast, the Docker configuration is
less memory-intensive than the VM configuration, leveraging the host
operating system's kernel. The \utnt configuration is the most
memory-efficient due to its specialization.

Next, we compared the configuration based on KSM. In this case, \utnt
(+KSM) exhibited a striking reduction in memory consumption, being
$17$ times less than Docker and an $74$ times less than the Vagrant
configuration. This outcome aligns with our expectations, as KSM
performs the merging of identical memory pages (after a certain period
of time), effectively reducing the memory footprint of the unikernel.
For Vagrant(+KSM), given its large memory consumption and the
volatility of its pages, KSM only manages to merge a tiny amount of
memory, which explains its low memory reduction.

\begin{figure}[!htp]
  \centering
  \includegraphics[width=0.95\linewidth]{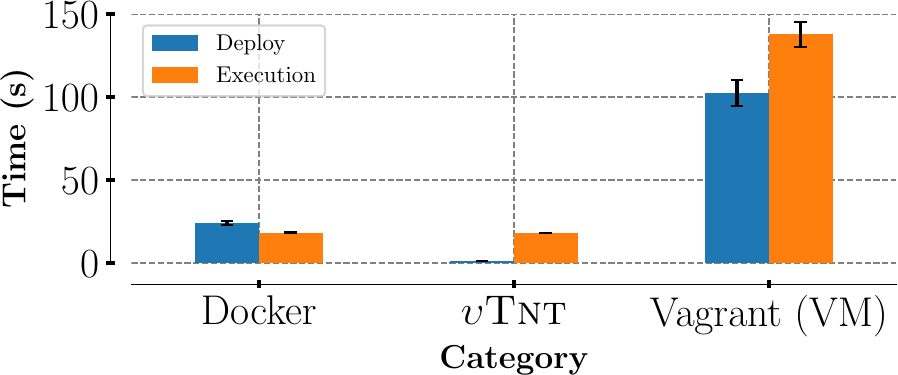}
  \caption{Total time (deployment + execution) to run \tnt, compared to
  Docker and Vagrant. Using \utnt gives the lower total time.}
  \label{evaluation.results.fig.total_execution_time}
\end{figure}

We then proceeded to evaluate the overall time, encompassing the
deployment and the execution time (creation, execution of \tnt with
specific parameters, instance destruction, and cleanup). Additionally,
we measured the CPU consumption associated with each deployment. To
avoid unnecessarily overloading the graphs for these two evaluations,
we did not consider KSM configurations, given that KSM operates on a
dedicated CPU core and has a negligible impact on the performance of
an individual instance running on another core. In this experiment, we
introduced a slight modification to the setup by incorporating a
deployment phase. During this phase, we considered a sandboxed
environment that necessitates initialization before deploying a
specific instance. For example, with Docker, a \texttt{docker pull}
operation is required to fetch the Docker image from a remote
repository. In the case of Vagrant, a \texttt{vagrant up} command is
necessary to launch the virtual machine. Lastly, for \utnt, it gets
the underlying image from a remote server. Once the environment has
been initiated, the image is executed, and the execution time is
measured. To ensure the statistical validity of the results, this
experiment was repeated 30 times. Results are shown in
Fig.~\ref{evaluation.results.fig.total_execution_time}.

We start by examining the \tnt execution time for each configuration.
Notably, the one for \utnt and Docker is relatively similar, as \utnt
involves initiating a unikernel from scratch, including to
initialize essential components such as the memory, the scheduler,
etc. In contrast, Vagrant's execution time is quite huge compared to
the two other configurations. This is explained by the necessity of
booting a heavyweight virtual machine. Consequently, this
discrepancy in execution time can significantly affect the \tnt
runtime, resulting in a duration that is up to $7.5$ times longer than
the one observed with \utnt and Docker.

The time it takes to deploy varies significantly based on the chosen
configuration. Deploying \utnt is the fastest option, requiring the
download of a small single-binary file (less than $1$~MB), taking only
a few milliseconds. Deploying with Docker is somewhat slower due to
the need to download a larger image (a few hundred megabytes) from
DockerHub~\cite{url_dockerhub}. Lastly, deploying with Vagrant is the
most time-consuming, as it involves downloading a virtual machine
image (a few gigabytes) from VagrantCloud~\cite{url_vagrantcloud}. All
in all, deploying and executing \utnt allows to have a gain of $2.4$
times compared to Docker and $13$ times compared to Vagrant.

\begin{figure}[!htp]
  \centering
  \includegraphics[width=0.95\linewidth]{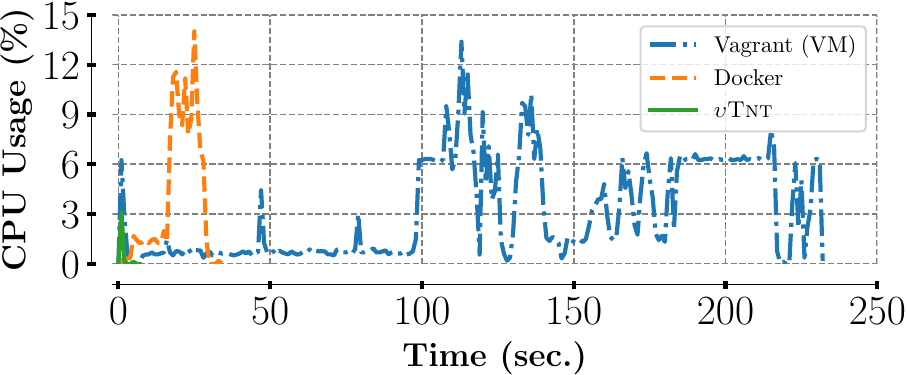}
  \caption{CPU consumption for \utnt, compared to \scamper
  implementation as a virtual machine (VM) and Docker.}
  \label{evaluation.results.fig.cpu}
\end{figure}

\begin{figure*}[!htp]
  \begin{center}
    \subfloat[Memory usage.]{
      \includegraphics[width=0.3\linewidth]{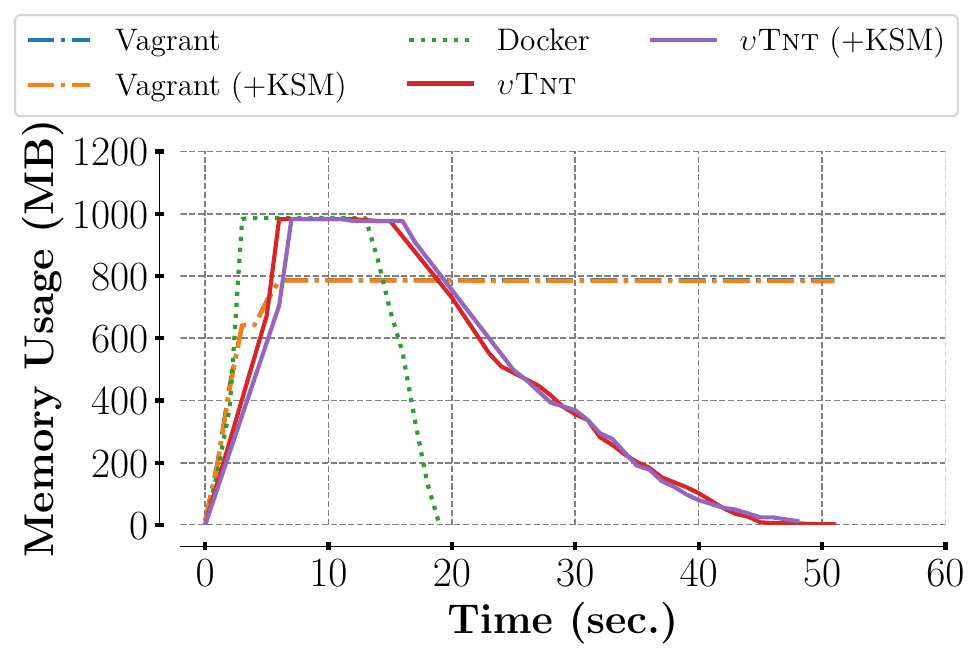}
      \label{evaluation.budget.memory} }
    \subfloat[CPU usage.]{
      \includegraphics[width=0.3\linewidth]{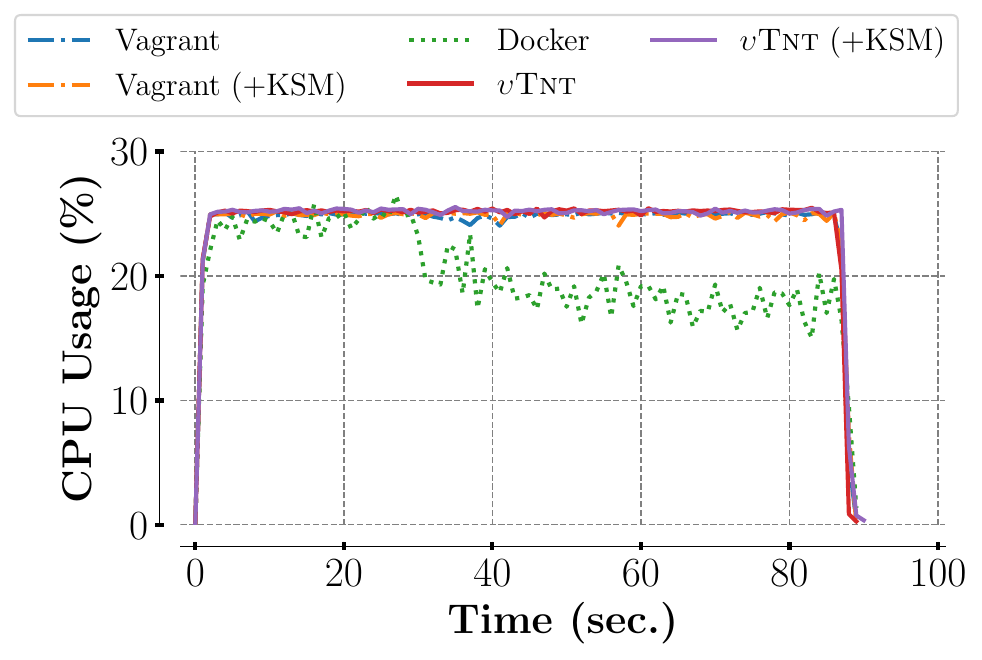}
      \label{evaluation.budget.cpu} }
    \subfloat[Number of instances.]{
      \includegraphics[width=0.3\linewidth]{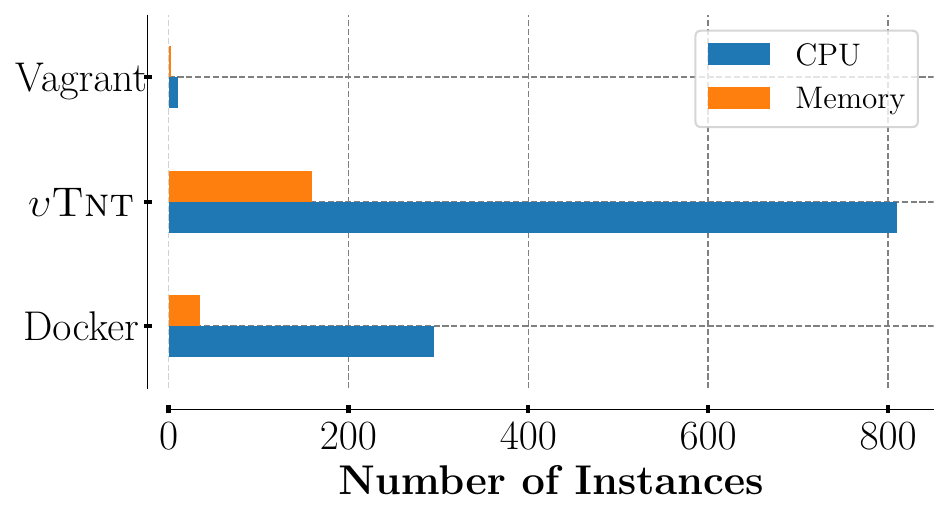}
      \label{evaluation.budget.instances} }
  \end{center}
  \caption{Memory and CPU usage with a fixed budget. Using \utnt allows to run considerably more instances than Docker and Vagrant.}
  \label{evaluation.budget}
\end{figure*}

From this same setup, we conducted CPU usage measurements for each
configuration. The findings, depicted in Fig.~\ref
{evaluation.results.fig.cpu}, reveal that \utnt exhibits lower CPU
utilization in comparison to Docker and Vagrant. This gap can be
attributed to several factors, including its minimalist design,
application-specific components, single-address space, effective I/O
management, reduced attack surface, and the utilization of a
lightweight hypervisor optimized for resource efficiency. In contrast,
Docker operates containers with a complete operating system, which
frequently leads to increased resource overhead. Vagrant is
considerably more expensive than \utnt and Docker. This is primarily
attributed to the fact that a full VM comes preloaded with a set of
services (e.g., systemd, cron, etc.), collectively driving up CPU
usage.

Concluding our study, we conducted two additional experiments in which
we initiated multiple instances of \utnt, Docker, and Vagrant, each
constrained by specific resource allocations: $1$~GB of memory and a
25\% CPU usage limit (4 cores). All these instances are started nearly
simultaneously, with a 0.1-second gap between each instance. In order
to prevent launching DOS attacks against the initial remote target, we
slightly updated the configurations by executing a \traceroute on the
address of the network interface and introduce a 10-second delay to
simulate a complete \traceroute. We chose arbitrary values for the
budget, which are representative of the resources that may be
available in a system. Nevertheless, it is expected that the outcomes
would be comparable even when different budget values are employed. 
Fig.~\ref{evaluation.budget} presents the
outcomes of these experiments. In every scenario, a noteworthy
observation emerges: \utnt consistently allows for a significantly
greater number of instances to be launched within the specified
resource constraints, outperforming both Docker and Vagrant.

Concerning the memory allocation, we employed Linux's control groups
to limit memory usage at $1$~GB across all configurations. As before,
we also considered configuration based on KSM. As can be seen in
Fig.~\ref{evaluation.budget}, using \utnt allows to obtain almost $5$
times more instances than Docker and $53$ times more instances than
Vagrant. In this experiment, it is important to note that the use of
KSM does not affect the number of instances. Indeed, given the
substantial number of pages to scan (for both Vagrant and \utnt), KSM
takes time to converge, making it difficult to reduce the initial
memory peak. This effect can be observed in
Fig.~\ref{evaluation.budget.memory}.

For the CPU usage, we relied on the \texttt{isolcpu} and
\texttt{taskset} commands to respectively isolate tasks and limit the
CPU usage to $25$\% (4 cores) for the different configurations. In
this experiment, we initiate multiple instances by pinning them to CPU
cores and allocating an equivalent execution time to every
configuration. When employing \utnt, this allows to launch almost $3$
times more instances than Docker and $81$ times more instances than
Vagrant. The CPU evolution of the different configurations is depicted
in Fig.~\ref{evaluation.budget.cpu}. In the context of Docker, we
observe decreased CPU utilization when compared to both \utnt and
Vagrant. This observation raises the possibility that Docker might be
performing more I/O operations. In that case, the CPU may spend a
significant amount of time waiting for I/O operations to complete,
leading to lower CPU utilization.

\section{Flexible Deployment}\label{usecase}

Our primary goal is to showcase the quick deployment of unikernels
for network measurements anywhere on the Internet. To achieve this, we
identify a relevant scenario: deploying unikernels on remote servers
for immediate network measurement purposes, where the deployment
process is orchestrated by a controller.

\subsection{Scenario}\label{usecase.perf.methodo}

In this experiment, we leveraged the OVH cloud
infrastructure~\cite{ovh_url} to provision five remote servers
situated in diverse locations, including Singapore, Canada, Australia,
France, and Poland. Each node has the same environment and
hardware. As before to prevent launching DOS attacks against the
initial remote target, we slightly updated the configurations by
executing \tnt on the address of the network interface. The objective
of the experiment is not to assess the network measurement but the
responsiveness and flexibility of different solutions. As before, we
repeated this experiment $30$ times, to ensure the statistical
validity of the results.

\begin{figure}[!t]
  \centering
  \includegraphics[width=0.7\linewidth]{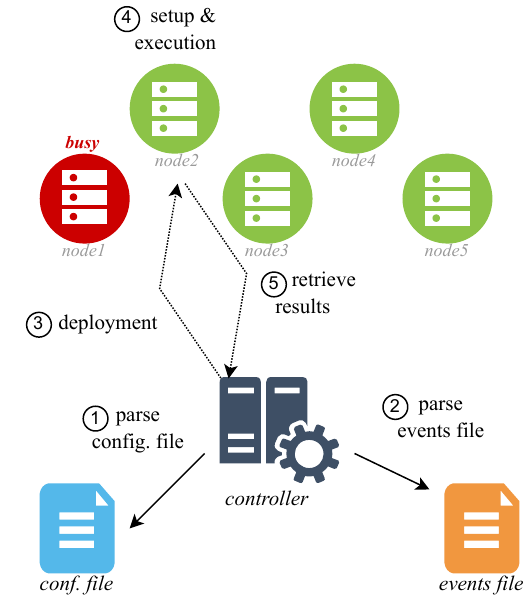}
  \caption{High-level overview of the controller. The controller
  deploys the configuration on a free node (\dfn{node1} being busy,
  processing a previous deployment).}
  \label{fig:diagram_deploy}
\end{figure}

We then developed a Python script to simulate a controller responsible
for deploying specific instances on remote servers for measurement
purposes. This script is executed on our local server, which has the
same hardware and configuration as previously described in Sec.~\ref
{evaluation.methodo}. To manage parallel deployments, we relied on a
multi-threaded architecture. As illustrated on Fig.~\ref
{fig:diagram_deploy}, the controller operates as follows: (1) It
reads a configuration file with remote node details.(2) Reads a list
of events for deployment timings (e.g., one deployment per second for
60 seconds). (3) Based on the list of events, the controller triggers
the deployment of the desired configuration (either \utnt, Docker, or
Vagrant) on the remote nodes. To accomplish this, the controller
sends a script. (4) When the script is received, the remote node
initializes its environment and then executes the specified
instance. (5) Remote node transmits execution results back to the
controller. The controller records cumulative time for each node, and
can adapt to high request scenarios with two distinct behaviours:
($i$) either it waits for a node to become available, ($ii$) it
ignores and discards the request.

Note that our approach is agnostic to the cloud provider and
can be used with other cloud providers such as Google
Cloud~\cite{url_google_cloud}, AWS~\cite{url_aws}, etc.

\subsection{Results}\label{usecase.perf.results}

\begin{figure}[!ht]
  \centering
  \includegraphics[width=0.95\linewidth]{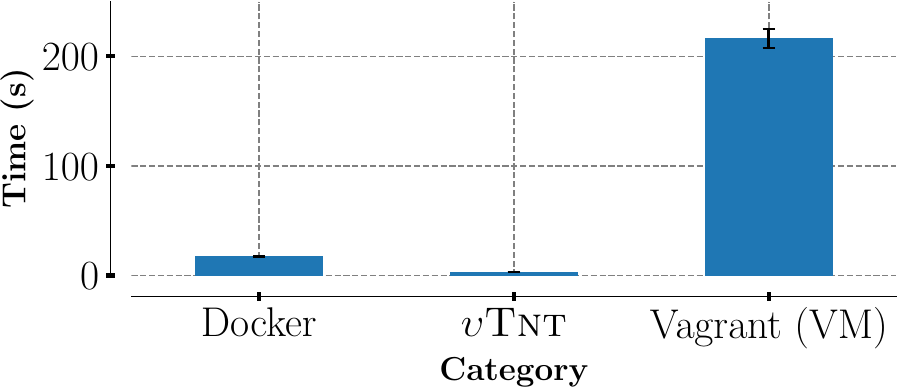}
  \caption{Average time to deploy and execute a specific configuration.
  \utnt is $5$ times faster than Docker and $65$ times faster than Vagrant.}
  \label{usecase.perf.results.fig.deploy_controller_time}
\end{figure}

We begin by discussing the total time (deployment + execution) of
various configurations. For this experiment, we focused on the first
controller's behaviour, where it waits for a node to become available
without discarding the request. As depicted in
Fig.~\ref{usecase.perf.results.fig.deploy_controller_time}, the \utnt
configuration exhibits the fastest performance, followed by Docker and
Vagrant. Using \utnt results in a deployment that is $5$ times faster
than Docker and $65$ times faster than Vagrant. This speed advantage
can be attributed to the fact that \utnt requires a shorter deployment
time compared to Docker and Vagrant, which are mainly hampered by the
size of the underlying image. The higher standard deviation observed
with Vagrant is likely attributed to the download of a relatively
large image which can vary according to the network and the node
location.

\begin{figure}[!ht]
  \centering
  \includegraphics[width=0.95\linewidth]{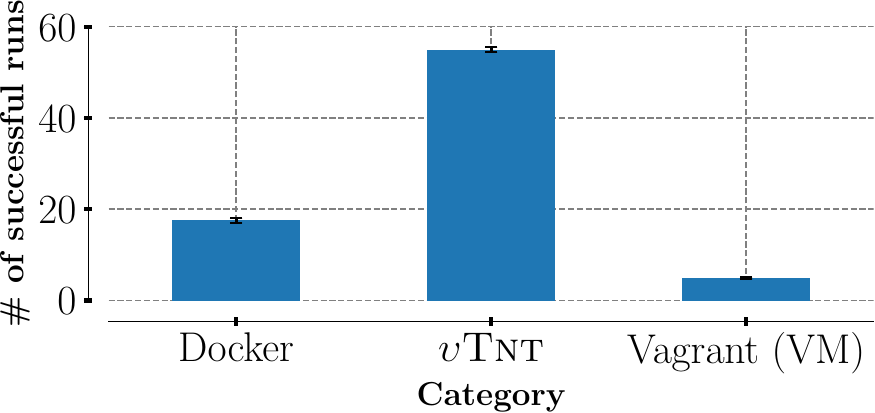}
  \caption{Number of successful deployments (+execution) per configuration.
  \utnt can achieve $3$ times more deployments than Docker and $11$
  times more than Vagrant.}
  \label{usecase.perf.results.fig.deploy_controller}
\end{figure}

Subsequently, we relied on the second controller's behaviour, which
involved ignoring and discarding a new deployment if none of the nodes
was available. We can observe the outcomes of this experiment in
Fig.~\ref{usecase.perf.results.fig.deploy_controller}. With this
approach, the controller accomplishes $90\%$ of \utnt deployments,
$28\%$ of Docker deployments, and $8\%$ of Vagrant deployments. Just
as previously noted, \utnt exhibits superior performance, followed by
Docker, while Vagrant demonstrates poor scalability.

\section{Discussion}\label{discussion}

\subsection{Porting to Unikraft}\label{discussion.porting}
For porting \tnt to Unikraft, we have chosen a manual approach, which
involves building the application from its source code. We made this
decision because the \scamper + \tnt codebase is not overly complex,
with few shared libraries and required primitives. Porting another
open-source measurement application would require following the same
steps and involve a similar porting effort as ours.

If an application is closed-source or exhibits complexity, such as
having numerous libraries, Unikraft offers an alternative approach
centered around binary compatibility. This means that unmodified Linux
Executable and Linkable Formats (ELFs) can be executed within
Unikraft. While this approach eliminates the need for manual porting
work, it does come with an initial cost of mapping system calls to the
underlying OS functions. However, there are two significant drawbacks:
the overhead resulting from multiple levels of indirection due to
system call compliance and the possibility of runtime crashes caused
by slight incompatibilities between the application's expected
ABI\footnote{\dfn{Application Binary Interface} (ABI) is a set of
rules defining how software components interact at the binary level,
ensuring compatibility and interoperability between different programs
and systems.} and the one provided by the underlying OS.

The choice between these two approaches ultimately lies with the developer
and maintainer.

\subsection{Cloud Deployment}\label{discussion.cloud}
In this paper, we have demonstrated that measurement infrastructures
could take advantage of unikernels to become more flexible and
efficient. However, in our evaluations, we only considered one
platform (KVM) and architecture (x86\_64). Unikernel frameworks come
with a set of supported platforms and
architectures~\cite{unikraft,nanos}. For instance, Unikraft can also
be deployed on a Xen server~\cite{xen_barham_2003} or be compiled to
run on an ARM architecture. We expect similar performance results on
these platforms.

\section{Conclusion}\label{ccl}
The last twenty years have seen the rise and development of measurement
infrastructures and efficient probing tools while, at the same time,
operating systems have evolved from monolithic architecture to
virtualization and specialization with unikernels. This paper embraced
those developments to introduce \utnt, a proof-of-concept of porting a
network probing tool (i.e., \tnt running over \scamper) in a
unikernel.

In this paper, we provided the complete toolchain when transforming
\tnt into \utnt. We believe this toolchain could be a very first step
towards the generalisation of unikernels in network infrastructures.
Indeed, this paper has demonstrated the supremacy of \utnt over more
traditional approaches. Further, this paper also discussed a case
study in which the flexibility and responsiveness of \utnt could find
a suitable usage, compared again to more traditional approaches.

\section*{Software Artifacts \& Acknowledgments}

All the code described and discussed in this paper (\utnt, performance
measurement, controller script, deployment over cloud provider) is
available on Gitlab~\cite{url_utnt}.

This is supported by the CyberExcellence project funded by the Walloon
Region, under number 2110186, and the Feder CyberGalaxia project.

{\balance
\bibliographystyle{IEEEtran}
\bibliography{Bibliography}
}

\end{document}